\documentclass[prb,aps,twocolumn,showpacs]{revtex4}
\usepackage{amsmath}
\usepackage{amssymb}
\usepackage{graphicx}
\usepackage{dcolumn}
\usepackage{graphicx}
\usepackage{dcolumn}
\usepackage{bm}

\begin{document}

\title{Magnetocapacitance oscillations and thermoelectric effect in two-dimensional
electron gas irradiated by microwaves.}

\author{A. D. Levin,$^1$ G. M. Gusev,$^1$  O. E. Raichev,$^2$ Z. S. Momtaz,$^1$
and A. K. Bakarov,$^{3,4}$}

\affiliation{$^1$Instituto de F\'{\i}sica da Universidade de S\~ao
Paulo, 135960-170, S\~ao Paulo, SP, Brazil}

\affiliation{$^2$Institute of Semiconductor Physics, NAS of
Ukraine, Prospekt Nauki 41, 03028 Kyiv, Ukraine}

\affiliation{$^3$Institute of Semiconductor Physics, Novosibirsk
630090, Russia}

\affiliation{$^4$Novosibirsk State University, Novosibirsk 630090,
Russia}

\date{\today}

\begin{abstract}

To study the influence of microwave irradiation on two-dimensional electrons, we apply
a method based on capacitance measurements in GaAs quantum well samples where the gate
covers a central part of the layer. We find that the capacitance oscillations at high magnetic
fields, caused by the oscillations of thermodynamic density of states, are not essentially
modified by microwaves. However, in the region of fields below 1 Tesla, we observe another set
of oscillation, with the period and the phase identical to those of microwave induced
resistance oscillations. The phenomenon of microwave induced capacitance oscillations
is explained in terms of violation of the Einstein relation between conductivity and
the diffusion coefficient in the presence of microwaves, which leads to a dependence of
the capacitor charging on the anomalous conductivity. We also observe microwave-induced
oscillations in the capacitive response to periodic variations of external heating. These
oscillations appear due to the thermoelectric effect and are in antiphase with microwave induced
resistance oscillations because of the Corbino-like geometry of our experimental setup.

\pacs{73.43.Qt, 73.50.Pz, 73.21.Fg, 73.50.Lw}

\end{abstract}

\maketitle

\section{Introduction}

The microwave (MW) irradiation of high-mobility two-dimensional (2D) electron gas in
the perpendicular magnetic field $B$ dramatically modifies the transport properties of the electron
system and leads to a remarkable phenomenon of microwave-induced resistance oscillations
(MIRO) and associated zero resistance states [1-3]. The periodicity of the oscillations
is determined by the ratio of the MW
frequency $\omega$ to the cyclotron frequency $\omega_c=|e|B/mc$ ($m$ is the effective
mass of the electron). Whereas the physics of MIRO is not fully understood, the majority
of experimentally observed features are described [3-9] by the theories based on
the consideration of scattering-assisted transitions of electrons between Landau levels
due to absorption and emission of MW radiation quanta. Besides, purely classical
approaches to magnetotransport [10 - 13] were also applied to the description of the
phenomenon, and it was suggested that MIRO might originate from the effects near
sample edges [11,12] and contacts [12] as opposed to the bulk transport properties.

The progress in understanding the effect of MW irradiation on 2D electrons in general
and the MIRO phenomenon in particular crucially depends on the information obtained from
experimental studies, which in turn relies on the diversity of applied experimental
methods. The basic method is a standard magnetotransport experiment, the measurements
of electrical resistance under external dc driving. The methods that do not use dc
driving are rare exceptions. They include measurements of the photovoltaic response
in the samples with built-in spatial variation of electron density near the contacts
[14 - 16], when the oscillating photovoltage appears as a result of MW irradiation,
and thermoelectric measurements [17], when the MW irradiation leads to oscillations
of thermoinduced (phonon-drag) voltage. In this paper, we propose and exploit another
method, based on capacitance measurements in the samples where the gate covers a central
part of the 2D layer.

Studies of the capacitive response commonly serve as a source of information about the
thermodynamic properties of 2D electron gas in magnetic field. In particular, measurements
of low-frequency electrical impedance in these systems allow one to find the thermodynamic
density of states $(\partial n/\partial \mu)_T$, where $n$ is the electron density and
$\mu$ is the chemical potential [18 - 23]. This is an efficient method to study the
density of states of electrons as well as the correlation effects, which is important for
understanding the integer and fractional quantum Hall effects. Recently, it was proposed
[24] to apply periodic variations of temperature $T$, induced by a heater to extract the
entropy density $(\partial S/\partial n)_T=-(\partial \mu/ \partial T)_n$, which gives complementary information about the properties of 2D electron systems.

Below we show experimentally that both voltaic and thermal perturbations in the presence
of MW irradiation lead to an oscillating capacitive response that bears a resemblance to MIRO.
A theoretical consideration gives an explanation of these effects, suggesting that the response
in both cases is directly related to MW-induced changes in electrical conductivity. In
spite of the fact that MW irradiation does not change the carrier density, it causes
redistribution of potentials in the 2D plane partly covered by the gate. We establish
two physical mechanisms of such influence. First, since the MW irradiation strongly modifies
the conductivity and does not change the diffusion coefficient, the Einstein relation becomes
violated [16]. This leads to a difference between electrochemical potentials in the regions under
the gate and out of the gate. Second, in the presence of temperature gradients, the capacitor
charging is influenced by the thermoinduced voltage between the Ohmic contact to 2D gas
and the heated region under the gate. Meanwhile, it is established that thermoinduced voltages are
modified by the MW irradiation [17,25], mostly because of the influence of the radiation on conductivity. It is worth to note that the MW-induced effects described above exist and are
actually the most prominent in the region of a relatively weak magnetic field, where the
Shubnikov-de Haas oscillations are thermally suppressed. In this region, the magnetic-field
induced changes in thermodynamic density of states and in entropy density are not seen
experimentally. Therefore, the oscillations of the capacitive response due to MW irradiation
are easily separable from the oscillations of thermodynamic quantities which appear at higher
magnetic fields.

The paper is organized as follows. In Section II we present the details of measurements and
experimental results. Section III contains the theoretical description, comparison of theory
with experiment, and discussion. Concluding remarks are given in the last section.

\section{Experiment}

We studied Schottky-gated narrow (14 nm) quantum wells (QW) with electron density
$n \simeq 6.5 \times 10^{11}$ cm$^{-2}$ and a mobility $\mu=2 \times 10^6$ cm$^2$/V s.
We used square ($8\times8$ mm$^{2}$) specimens of Van der Pauw geometry, with
contacts at the corners. A 4.5 nm TiAu semitransparent gate of nearly circular shape
with area $A=4$ mm$^{2}$ was evaporated on top of the sample. The distance between
the QW and the sample surface was $d_G=115$ nm, and the gate-QW geometric capacitance
$C_{0}$ was 3.9 nF.
The capacitance was measured by the modulation technique: the gate voltage was modulated
with a small ac voltage of 10-50 mV at frequencies in the range $1.5 -3.1$ Hz and the
recharging current was measured using a lock-in amplifier.
The geometric capacitance and the frequencies were small enough to neglect contribution
of the resistance of the 2D electron gas in the recharging current signal.
We also measured low-frequency ac voltage proportional to the recharging current
generated due to temperature modulation by applying a technique similar to that used in
Ref. 24. The heater was placed on the back side of the sample in order to minimize the temperature
gradient over the sample area. The heater voltage was modulated by the law $v_{0}+v_{0} \cos(
\omega_0 t)$, with $f_0=\omega_0/2 \pi \approx 3.24$ Hz, leading to
modulation of heating
power with the same frequency. The corresponding variation of electron temperature $\Delta T \sim
0.2$ K was estimated from the amplitude of the Shubnikov-de
Haas oscillations. The measurements were carried out in a VTI cryostat with a waveguide to
deliver microwave (MW) irradiation (frequency range 110 to 170 GHz) down to the sample. Several
devices from the same wafer were studied.

\begin{figure}[ht]
\includegraphics[width=9cm]{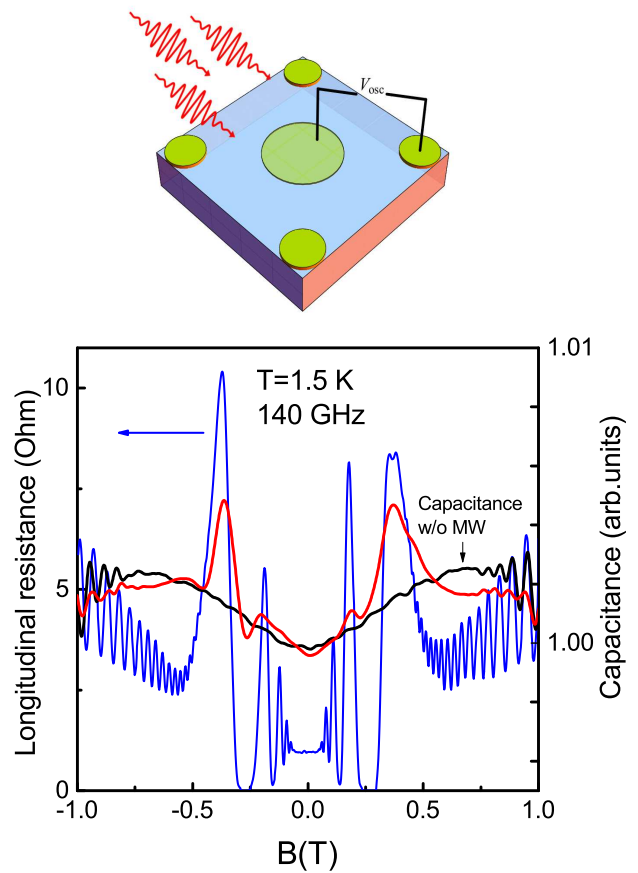}
\caption{Magnetoresistance and magnetocapacitance under microwave irradiation of frequency 140 GHz.
Dark magnetocapacitance is also shown. The top part shows the geometry of the experiment.}
\end{figure}

\begin{figure}[ht]
\includegraphics[width=9cm]{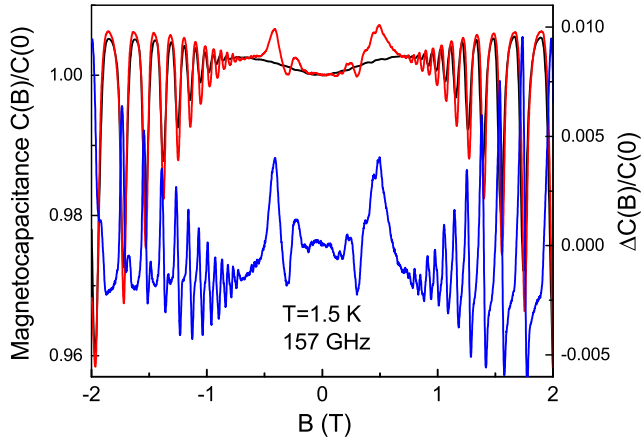}
\caption{Magnetocapacitance, dark and under MW irradiation at 157 GHz, in a wide interval
of magnetic fields. The lower plot shows the difference $\Delta C$ between the MW-induced
and dark magnetocapacitances.}
\end{figure}

In Fig. 1 we present the longitudinal magnetoresistance under microwave irradiation for 140 GHz
and at a temperature of 1.5 K. The resistance reveals strong MIRO and zero resistance state
evolved from the last minimum. This figure also shows typical magnetocapacitance
(capacitance $C(B)$ normalized to its zero-field value) traces
both with MW and without it. The presence of MW irradiation results in an oscillating
magnetocapacitance contribution with much weaker relative amplitude in comparison with
phototoresistance. The period and the phase of the photocapacitance oscillations
coincide with those of MIRO, though we do not see any indications of a zero resistance state
regime in the main minimum of the capacitance. In Fig. 2 we plot the magnetocapacitance for
an extended interval of $B$ in order to show the pronounced oscillations caused by consecutive
passage of Landau levels through the Fermi level when the Landau levels become separated
in strong magnetic fields. The MW irradiation does not lead to a qualitative change of the
shape of these oscillations, though the amplitudes slightly increase. We point out that
this behavior is not usual, as one should expect some decrease in the amplitudes because
of heating of the electron gas by microwaves.

\begin{figure}[ht]
\includegraphics[width=9cm]{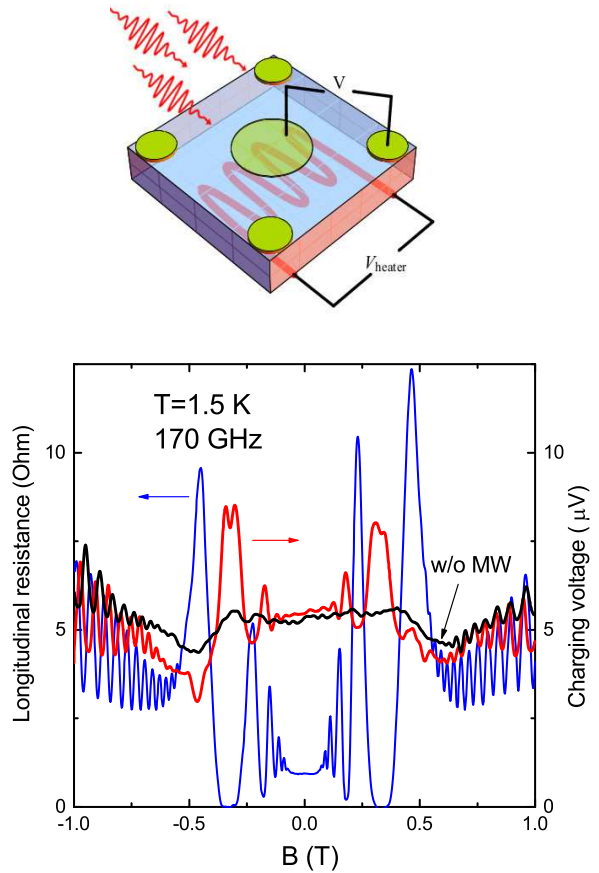}
\caption{Magnetoresistance and charging voltage under microwave irradiation of frequency 170 GHz.
Dark charging voltage is also shown. The top part shows the geometry of the experiment.}
\end{figure}

Figure 3 illustrates magnetooscillations of recharging voltage induced by temperature modulation
under MW irradiation in comparison with MIRO. The periodicity of these oscillations is the
same as that of MIRO. However, in contrast to MW-induced magnetocapacitance oscillations,
the oscillations of thermoinduced voltage have a much larger amplitude and occur in
antiphase with MIRO. These properties have been checked in the measurements for several different
MW frequencies. The charging voltage without MW irradiation (dark signal) demonstrates
weak features identified as the magnetophonon oscillations due to resonance scattering
of electrons by acoustic phonons. These oscillations were observed previously in phonon-drag thermopower measurements [17,26] as well as in the magnetoresistance of 2D
electron systems (see Ref. 3 for a review). The presence of magnetophonon oscillations
suggests that the charging is influenced by thermoelectric effects mediated by phonon drag.
The oscillating contribution added by MW irradiation is much larger than the magnetophonon
oscillations.

\section{Theory and discussion}

The experimental data presented above show that microwaves have a non-trivial effect on
gate-QW capacitor charging. This effect resembles the influence of microwaves on resistance
and depends on the type of excitation source, whether electric or thermal. The theory
given in this section provides a detailed account of the processes contributing to the
observed capacitive response.

The recharging current is given by the time derivative of the induced electric charge
$\delta Q$:
\begin{equation}
J_t=\frac{d \delta Q_t}{dt}=e \int_G d {\bf r} \frac{d n_{{\bf r}t}}{dt},
\end{equation}
where $n_{{\bf r}t}$ is the local electron density and the integral is taken over the
region covered by the gate. Below, the time index is dropped out because the recharging
occurs in the quasi-stationary regime. The variation of the electrostatic potential in the 2D
layer under the gate is related to the variation of the electron density as $\delta \phi=
(4 \pi e d_G/\epsilon_0) \delta n$, where $\epsilon_0$ is the dielectric permittivity
of the cap layer and $d_G$ is the distance between the gate and the 2D layer. Since the
electrochemical potential $\eta$, which determines the local voltage $V_{\bf r}=
\eta_{\bf r}/e$, is connected with electrostatic potential and chemical potential
by the relation
\begin{equation}
\eta_{\bf r}=e\phi_{\bf r}+\mu_{\bf r},
\end{equation}
the variation of the voltage under the gate is written as
\begin{equation}
e \delta V=\delta \mu + (4 \pi e^2 d_G/\epsilon_0) \delta n.
\end{equation}
In this expression, the coordinate index is dropped out because we consider the
quantities averaged over the areas of the size exceeding all microscopic lengths
including the screening length, so $\delta n$ is assumed to be coordinate-independent.
The local equilibrium condition (the absence of currents) requires $\delta V$ to
be coordinate-independent if the temperature under the gate is uniform, which is also
assumed. Equation (3), with the use of the relation
$$
\delta \mu =\frac{\partial \mu}{\partial n} \delta n + \frac{\partial \mu}{\partial T} \delta T
$$
leads to electric charge variation as a response to both the voltage and the temperature variations:
\begin{equation}
\delta Q = C [\delta V - e^{-1}(\partial \mu/\partial T) \delta T],
\end{equation}
with the capacitance
\begin{equation}
C=C_0 \left[1 + \frac{C_0}{e^2A} \left(\frac{\partial n}{\partial \mu} \right)^{-1} \right]^{-1},
\end{equation}
where $A$ is the area under the gate and $C_0 \simeq \epsilon_0 A/4 \pi d_G$ is the geometric capacitance.

Therefore, the capacitive response to perturbations $\delta V$ and $\delta T$ is described by
the isothermic derivative $\partial n/\partial \mu$ and the derivative $\partial \mu/\partial T$
at constant electron density. For non-interacting electron gas in thermodynamic equilibrium,
these quantities are entirely described by the density of states in the magnetic field:
\begin{equation}
\frac{\partial n}{\partial \mu} = \rho_{2D} \int d \varepsilon \left(-\frac{\partial f_{\varepsilon}}{\partial \varepsilon} \right) D_{\varepsilon}
\end{equation}
and
\begin{equation}
\frac{\partial \mu}{\partial T} =- \rho_{2D} \left(\frac{\partial n}{\partial \mu} \right)^{-1}
\int d \varepsilon \left(-\frac{\partial f_{\varepsilon}}{\partial \varepsilon} \right) D_{\varepsilon} \frac{\varepsilon-\mu}{T},
\end{equation}
where $D_{\varepsilon}$ is the density of states expressed in units $\rho_{2D}=m/\pi \hbar^2$
and $f_{\varepsilon}$ is the equilibrium distribution
function. If the density of states is represented as a series of oscillating
harmonics, $D_{\varepsilon}=1+2\sum_{k=1}^{\infty} a_k
\cos(2 \pi k \varepsilon/\hbar \omega_c)$, these expressions are rewritten as
\begin{equation}
\frac{\partial n}{\partial \mu} = \rho_{2D} {\cal S}_0, ~~\frac{\partial \mu}{\partial T}
=-\frac{{\cal S}_1}{ {\cal S}_0},
\end{equation}
\begin{eqnarray}
{\cal S}_0=1+2\sum_{k=1}^{\infty} a_k \frac{X_k}{\sinh X_k}
\cos \left(\frac{2 \pi k \mu}{\hbar \omega_c} \right), \nonumber \\
{\cal S}_1= 2 \pi \sum_{k=1}^{\infty} a_k \frac{\partial}{\partial X_k}
\frac{X_k}{\sinh X_k}
\sin \left(\frac{2 \pi k \mu}{\hbar \omega_c} \right),
\end{eqnarray}
where $X_k=2 \pi^2 k T/\hbar \omega_c$. Equation (9) demonstrates the presence of quantum
oscillations similar to Shubnikov-de Haas oscillations. Such kind of oscillations are known
to be suppressed because of random spatial fluctuations of the chemical potential [27].
If such variations are described by a Gaussian distribution $\propto \exp[-(\mu-\overline{\mu})^2/\Gamma^2]$,
their influence can be taken into account
by multiplying the terms under the sums in Eq. (9) by the factors
$\exp[-(\pi \Gamma k/\hbar \omega_c)^2]$.

Let us discuss in which way the microwaves can modify the quantities $\partial n/\partial
\mu$ and $\partial \mu/\partial T$. First of all, the MW irradiation leads to some heating
of electron gas and, therefore, suppresses the oscillations as factors $X_k$ increase.
This is, however, a trivial effect which we are not interested in below. A very intensive
radiation may modify the electron spectrum, but we did not work with such powerful MW sources
in our experiment. Also, the MW irradiation creates an additional oscillating part of the
distribution function, $\delta f_{\varepsilon}^{(\omega)}$, which is responsible for the
"inelastic" mechanism [7] of MIRO. However, since the radiation does not change the number
of electrons in the system, one has the identity $\int d \varepsilon D_{\varepsilon} \delta f_{\varepsilon}^{(\omega)}=0$ which means that both $\partial n/\partial \mu$ and
$\partial n/\partial T$ remain unchanged. The quantity $\partial \mu/\partial T$ may change,
but the relative contribution of this effect is estimated, from a direct calculation,
as $\hbar \omega/\mu$ times smaller than the relative effect of microwaves on the
conductivity. Therefore, we conclude that the influence of MW irradiation on the
capacitive response through the quantities $\partial n/\partial \mu$ and $\partial
\mu/\partial T$ is basically reduced to the effect of heating of the electron gas. Another
kind of influence, through the transport properties of the 2D layer, is more essential and
considered below.

The response of the system to external perturbations is determined not only by the
intrinsic properties of the capacitor between the 2D gas and the gate, but also by
the potential distribution in the 2D plane. In other words, knowing the relation Eq. (4)
between $\delta Q$, $\delta V$, and $\delta T$ is not sufficient to solve the problem
since $V$ is the voltage under the gate, while in our sample the contact to the 2D gas is
placed outside the gate and stays at a different voltage $V_0$. To find a relation between
$V$ and $V_0$, we present the local current density as
\begin{equation}
{\bf j}_{\bf r}={\bf j}^{(0)}_{\bf r}+{\bf j}^{(t)}_{\bf r}+{\bf j}^{(\omega)}_{\bf r},
\end{equation}
where ${\bf j}^{(0)}_{\bf r}$ is the current in the absence of microwave
irradiation and thermal excitation, ${\bf j}^{(t)}_{\bf r}$ is the thermoinduced current,
and ${\bf j}^{(\omega)}_{\bf r}$ is the microwave-induced current. In the linear
response regime, ${\bf j}^{(0)}_{\bf r}$ is proportional to the gradient of electrochemical
potential $\eta_{\bf r}$ according to ${\bf j}^{(0)}_{\bf r}=-\hat{\sigma}^{(0)}
\nabla \eta_{\bf r}/e$, where $\hat{\sigma}^{(0)}=\sigma_d-\hat{\epsilon}\sigma_H$
is the conductivity tensor in magnetic field. In particular, $\sigma_H=e^2n/m\omega_c$
is the Hall conductivity, $\hat{\epsilon}$ is the $2 \times 2$ Levi-Civita matrix,
\begin{equation}
\sigma_d=\sigma_0 \int d \varepsilon \left(-\frac{\partial f_{\varepsilon}}{\partial \varepsilon} \right) D^2_{\varepsilon}
\end{equation}
is the longitudinal (dissipative) conductivity, and $\sigma_0=e^2n/m\omega^2_c \tau_{tr}$
is its classical part expressed through the transport time $\tau_{tr}$. We consider the
case of a classically strong magnetic field, $\omega_c \tau_{tr} \gg 1$, relevant for
high-mobility electron gas. If $\eta$ is constant, the current ${\bf j}^{(0)}_{\bf r}$
is zero.

The thermoinduced current in GaAs quantum wells is dominated by phonon drag contribution.
In general, this current is modified in the presence of microwave irradiation [25], but the
effect is small compared to the influence of microwaves on the resistivity and will be neglected
in the following. Using the model of bulk phonons with a three-dimensional wave vector
${\bf Q}=({\bf q}, q_z)$ and taking into account that the scattering of electrons by phonons is
quasi-elastic because of the smallness of the phonon energies compared to Fermi energy, one gets [25]
\begin{eqnarray}
{\bf j}^{(t)}_{\bf r}
=\frac{|e| k_F^2 m}{4 \pi^3 \hbar^4 \omega_c^2}  \int_0^{2 \pi}d\varphi \int_0^{\pi}
d\zeta \sin^{-2} \zeta
 \nonumber \\
\times \sum_{\lambda} \int_{0}^{\pi} \frac{d \theta}{\pi}  (1-\cos \theta) C_{\lambda {\bf Q}} I_{q_z} N_{\lambda {\bf Q}}({\bf r}) \int d \varepsilon
D_{\varepsilon} ~~~~ \\
\times \sum_{l=\pm 1} l D_{\varepsilon-l \hbar \omega_{\lambda {\bf Q}}}
(f_{\varepsilon-l \hbar \omega_{\lambda {\bf Q}}} - f_{\varepsilon} )
[\tau_{tr}^{-1} D_{\varepsilon} {\bf n}_{\varphi} - \omega_c \hat{\epsilon} {\bf n}_{\varphi}],
\nonumber
\end{eqnarray}
where $k_F=\sqrt{2 \pi n}$ is the Fermi wavenumber, $\lambda$ is the mode index,
$\varphi$ is the polar angle of ${\bf q}$, $\zeta$ is the
inclination angle, and $\theta$ is the scattering angle. The phonon frequency is given by
$\omega_{\lambda {\bf Q}}=s_{\lambda} \sqrt{q^2+q_z^2}$, where $q=2k_F\sin(\theta/2)$ and
$q_z=q \cot \zeta$. The function $C_{\lambda {\bf Q}}$ is the squared matrix element of the
electron-phonon interaction in the bulk, comprising deformation-potential and piezoelectric-potential
mechanisms (the detailed expression can be found, for example, in Ref. 25). The squared
overlap integral $I_{q_z}$ is defined as $I_{q_z}=| \int dz e^{iq_z z} F^2(z)|^2$, where $F(z)$
is the ground-state wavefunction describing confinement of electrons in the quantum well. Next,
${\bf n}_{\varphi}=(\cos \varphi,\sin \varphi)$ is a unit vector along ${\bf q}$. The current
${\bf j}^{(t)}_{\bf r}$ is determined by the distribution function of phonons, $N_{\lambda
{\bf Q}}({\bf r})$, which depends on the heater temperature $T_h$ and on all the details
of the sample geometry and composition. In spite of the fact that this function is unknown,
Eq. (12) can be rewritten in the general linear form
\begin{eqnarray}
{\bf j}^{(t)}_{\bf r}= \hat{{\cal B}} {\bf G}_{\bf r},~~~ {\bf G}_{\bf r}=-\nabla {\cal T}_{\bf r},
\end{eqnarray}
where ${\bf G}$ is a potential field proportional to the frictional drag force inflicted by
the in-plane component of phonon flux and ${\cal T}$ is a scalar potential associated with
this field. The symmetry properties of the tensor $\hat{{\cal B}}$ are the same as those of the conductivity tensor: $\hat{{\cal B}}={\cal B}_d-\hat{\epsilon} {\cal B}_H$, ${\cal B}_d$ is
even in $B$, ${\cal B}_H$ is odd in $B$, and $|{\cal B}_H | \gg |{\cal B}_d|$ under the assumed condition of a classically strong magnetic field. Under certain approximations, when $N_{\lambda
{\bf Q}}({\bf r}) \propto {\bf q} \cdot \nabla T$, one may identify ${\cal T}$ with temperature
$T$. In this case, $\hat{{\cal B}}$ is the thermoelectric tensor $\hat{\beta}$; see Ref. 25
for the theory of phonon-drag thermoelectric effect in a quantizing magnetic field.
Both ${\cal B}_d$ and ${\cal B}_H$ contain classical (non-oscillating) and
quantum (oscillating with $B$) contributions, the latter include the magnetophonon oscillations.

Finally, the microwave irradiation adds a photocurrent whose most essential part is
proportional to the gradient of electrostatic potential [3], [16]. At low temperatures
of 1-2 K, when the inelastic mechanism [7] of MW-induced photocurrent dominates,
the whole photocurrent is proportional to $\nabla \phi$, so we write
\begin{eqnarray}
{\bf j}^{(\omega)}_{\bf r}=-\sigma_{\omega} \nabla \phi_{\bf r}.
\end{eqnarray}
Since the term ${\bf j}^{(\omega)}_{\bf r}$ leads to violation of the Einstein relation
between conductivity and the diffusion coefficient, the quantity $\sigma_{\omega}$ has been
named by the authors of Ref. 16 as anomalous conductivity. In the case of weak excitation,
when the response is linear in MW power, the anomalous conductivity is given by the expression
\begin{eqnarray}
\sigma_{\omega}=\sigma_0 \frac{\tau_{in}}{4\tau_{tr}} P_{\omega} \int d \varepsilon
D^2_{\varepsilon} \sum_{l=\pm 1} \frac{\partial D_{\varepsilon+ l \hbar\omega}}{\partial \varepsilon}
(f_{\varepsilon+ l \hbar \omega}-f_{\varepsilon} ),
\end{eqnarray}
where $\tau_{in}$ is the inelastic scattering time, $P_{\omega}= (2 e^2 E^2_{\omega}
\mu/m \hbar^2 \omega^2) (|s_+|^2+|s_-|^2)$ is a dimensionless function proportional
to the MW power absorbed by 2D electrons, $E_{\omega}$ is the MW electric field, $s_{\pm}=({\rm e}_x
\pm i {\rm e}_y)/[\sqrt{2}(\omega \pm \omega_c + i \omega_p)]$, ${\bf e}$ is the unit vector of MW
polarization, and $\omega_p=4 \pi e^2 n/[ m c (1+\sqrt{\epsilon_0})]$ is the radiative decay rate.
In the systems with uniform electron density and temperature, where the chemical potential
is constant, one has $e \nabla \phi_{\bf r}=\nabla \eta_{\bf r}$, so the conductivity
$\sigma_{\omega}$ is simply added to $\sigma_d$ and the response is determined by the total
MW-modified dissipative conductivity $\sigma_d+\sigma_{\omega}$. If the chemical potential
depends on coordinate, the response depends separately on $\sigma_d$ and $\sigma_{\omega}$.

Since no current is injected into the 2D system through the contacts, one may write
\begin{equation}
\oint d {\bf l} \cdot \hat{\epsilon} {\bf j}_{\bf r} =0.
\end{equation}
where the integral is taken along any closed contour in the 2D plane. Equation (16) is the integral
form of the continuity equation $\nabla \cdot {\bf j}_{\bf r} =0$. It means that the total current
flowing in or out of the area encircled by a closed contour is zero. With the use of
Eq. (2), expressions for the currents, and the identity $\oint d {\bf l} \cdot \nabla \eta_{\bf
r}=0$ providing continuity of $\eta_{\bf r}$, one may rewrite Eq. (16) as
\begin{eqnarray}
\oint d {\bf l} \cdot \hat{\epsilon} \biggl[ (\sigma_d+\sigma_{\omega}) \nabla \eta_{\bf r}  \\
 - \sigma_{\omega} \left( \frac{\partial \mu}{\partial n} \nabla n_{\bf r} + \frac{\partial \mu}{\partial T}
\nabla T_{\bf r} \right) - e {\bf j}^{(t)}_{\bf r} \biggr]=0. \nonumber
\end{eqnarray}
In our experimental setup, the electron density $n$ changes in a narrow, determined by
the distance $d_G$, region near the edge of the gate. In the presence of microwaves, this causes a
corresponding jump of the electrochemical potential. To show this, we consider two closely spaced
contours congruent with the shape of the gate near the gate edge, a contour under the gate
and another contour ${\cal C}_1$ outside the gate, and integrate Eq. (17) between these contours.
Neglecting a possible change of temperature and contribution of thermoinduced currents in the narrow
interval of integration, we obtain
\begin{equation}
e (V-V_1)= \frac{\sigma_{\omega}}{\sigma_d+\sigma_{\omega}}
\left(\frac{\partial n}{\partial \mu} \right)^{-1} \delta n,
\end{equation}
where $V_1$ is the voltage outside the gate in the near-gate region, formally obtained by
averaging of the electrochemical potential over the contour ${\cal C}_1$ (see Fig. 4). The fact
that Eq. (18) does not contain the Hall conductivity is related to the Corbino-like geometry of
the experiment. However, in the presence of microwaves, when $V \neq V_1$, a large
non-dissipative Hall current proportional to $\sigma_{H}$ circulates along the
gate edge.

\begin{figure}[ht]
\includegraphics[width=9cm]{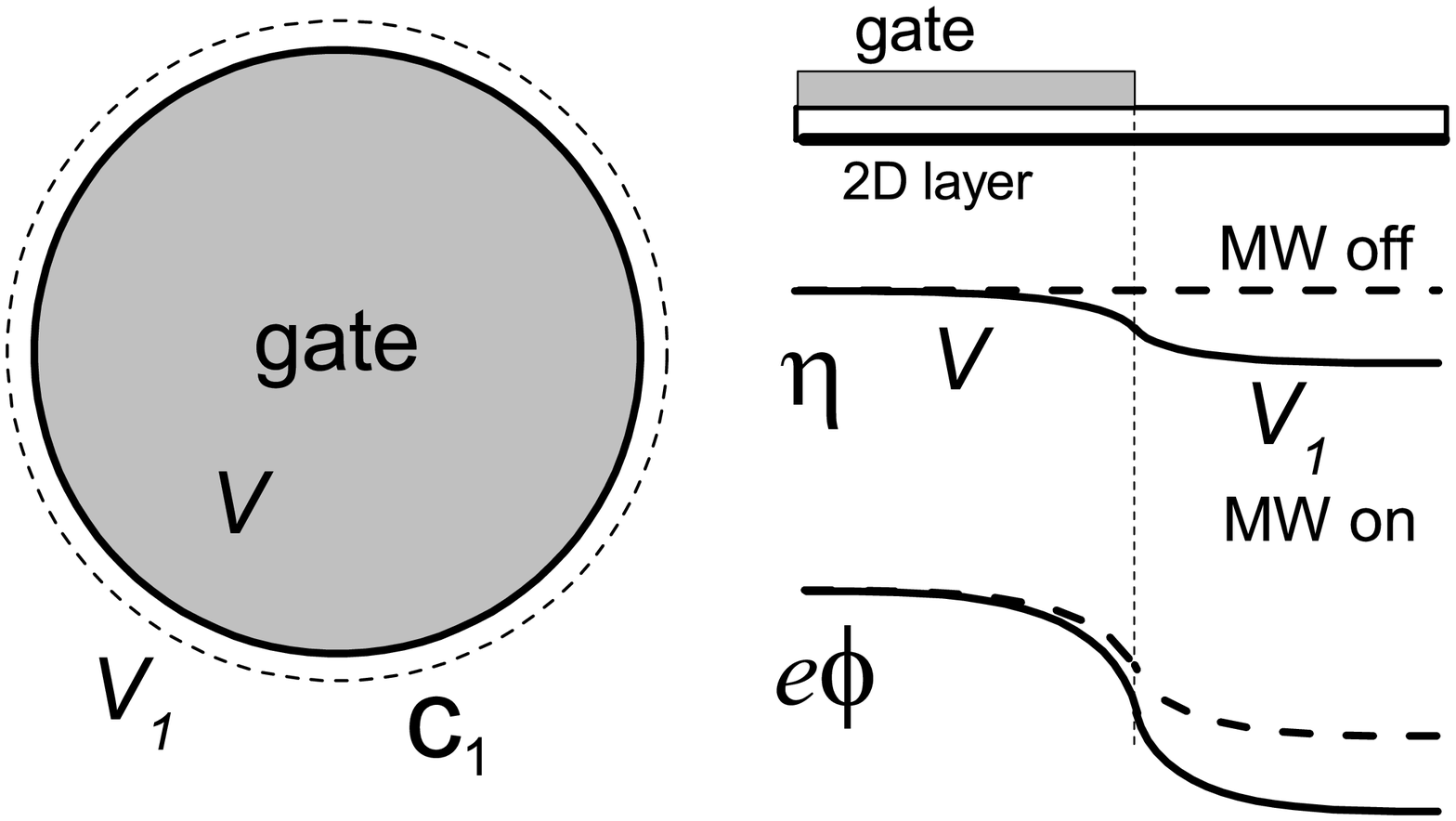}
\caption{Charging in the presence of microwaves. Left: schematic top view, right:
spatial dependence of electrochemical and electrostatic potentials. When the gate
is biased, the electron density under the gate becomes different from the density
outside the gate, and the electrostatic potential is changed near the gate edge. The
electrochemical potential is constant since the current must be zero. When the
sample is irradiated by microwaves, the condition of zero dissipative current
requires a change of electrochemical potential near the gate edge.}
\end{figure}

In the absence of non-equilibrium phonon fluxes and temperature gradients,
the electrochemical potentials are equal to $V$ and $V_1$ in the whole regions under
the gate and outside the gate, respectively. Consequently, the variation of the voltage
at the contact, $\delta V_0$, is equal to the variation of the voltage outside the gate,
$\delta V_1$, and we obtain
\begin{equation}
\delta Q=C_{MW} \delta V_0
\end{equation}
with the MW-modified capacitance
\begin{equation}
C_{MW}=C_0 \left[1 + \frac{\sigma_d}{\sigma_d+\sigma_{\omega}} \frac{C_0}{e^2A}
\left(\frac{\partial n}{\partial \mu} \right)^{-1} \right]^{-1}.
\end{equation}

To solve the problem in the presence of external heating, we notice that since the heater
area extends over the gate region, one can expect nearly uniform heating and non-essential
phonon drag in the gate region. This means that the electrochemical potential under the
gate is still uniform and equal to $eV$. In the region out of the gate, the electrochemical
potential changes because of the thermoelectric effect. This potential can be found from the
continuity equation $\nabla \cdot {\bf j}_{\bf r}=0$ with boundary conditions expressing
zero current normal to the sample boundary at this boundary and the equality
$\eta_{\bf r}=eV_1$ near the gate edge, ${\bf r} \in {\cal C}_1$. As the current is
given by the expression
\begin{eqnarray}
{\bf j}_{\bf r}= -\frac{1}{e} [\sigma_d+\sigma_{\omega} -  \sigma_H \hat{\epsilon}] \nabla \eta_{\bf r}
+\sigma_{\omega} \frac{1}{e} \frac{\partial \mu}{\partial T} \nabla T_{\bf r} \nonumber \\
- ( {\cal B}_d - {\cal B}_H \hat{\epsilon}) \nabla {\cal T}_{\bf r},
\end{eqnarray}
the continuity equation becomes the Poisson-like equation
$(\sigma_d+\sigma_{\omega}) \nabla^2 \eta_{\bf r}=\sigma_{\omega} (\partial \mu/\partial T) \nabla^2 T_{\bf r}-e {\cal B}_d \nabla^2 {\cal T}_{\bf r} \equiv \nabla^2 \Phi_{\bf r}$. This
equation has an obvious solution (here $\Phi_1$ is $\Phi_{\bf r}$ at ${\bf r} \in {\cal C}_1$):
\begin{eqnarray}
\eta_{\bf r}=eV_1+\frac{\Phi_{\bf r}-\Phi_{1}}{\sigma_d+\sigma_{\omega}}
\end{eqnarray}
which satisfies the boundary conditions in a particular case when both the sample
and the heater are radially symmetric (Corbino geometry [28 - 30]) and describes the
distribution with zero azimuthal gradients of $\eta_{\bf r}$ and zero radial current
in any point of the sample. Despite a different geometry of our sample (Fig. 5),
the solution (22) still remains valid at the symmetry axis $Ox$ because the symmetry requires
$[\nabla_y \eta_{\bf r}]_{y=0}=0$. Therefore, by using Eq. (22) one can calculate
the voltage difference $V_2-V_1$, where $V_2$ is the voltage at the boundary point
${\bf r}=(L/2,0)$ (see Fig. 5). To find the bias $V_0-V_2$, we write the boundary
condition for the $x$ component of the current at $x=L/2$: $j^x_{\bf r}|_{x=L/2}=0$.
In the case of classically strong magnetic fields, the main contribution to the
current comes from its Hall component, and the boundary condition is approximately
rewritten as $[\sigma_H \nabla_y \eta_{\bf r} + e {\cal B}_H \nabla_y {\cal T}_{\bf r}
]_{x=L/2} \simeq 0$. This leads to the relation $V_0-V_2=- ({\cal B}_H/\sigma_H)({\cal
T}_0-{\cal T}_2)$, where ${\cal T}_0$ and ${\cal T}_2$ correspond to the boundary points
$(L/2,L/2)$ and $(L/2,0)$ with the voltages $V_0$ and $V_2$, respectively.

\begin{figure}[ht]
\includegraphics[width=9cm]{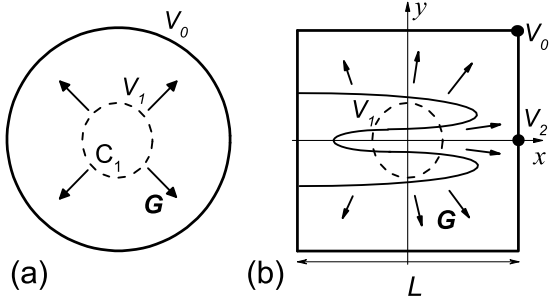}
\caption{Thermoelectric effect in the simple Corbino geometry (a) and in the
actual geometry of the sample (b). The arrows show the distribution of phonon drag
forces and temperature gradients. The meander in (b) schematically represents
the heater.}
\end{figure}

In summary,
\begin{eqnarray}
V_{10} \equiv V_1-V_0 \simeq -\frac{{\cal B}_H({\cal T}_2-{\cal T}_0)}{\sigma_H} \nonumber \\
-\frac{{\cal B}_d({\cal T}_1-{\cal T}_2)}{\sigma_d+\sigma_{\omega}} +
\frac{1}{e} \frac{\sigma_{\omega}}{\sigma_d+\sigma_{\omega}} \frac{\partial \mu}{\partial T}
(T_1-T_2).
\end{eqnarray}
Although this expression contains a number of unknown parameters, it gives a complete
account for the dependence of thermoinduced voltage on the MW-induced conductance
contribution $\sigma_{\omega}$. Finally, a unified expression relating the charging
$\delta Q$ with the external perturbations is written in the following way
\begin{equation}
\delta Q=C_{MW} \left[\delta V_0  + \delta V_{10}(T_h)- \frac{1}{e} \frac{\partial \mu}{\partial T} \delta T(T_h) \right],
\end{equation}
where we have emphasized that the variations of the temperature of electron gas under the
gate, $\delta T$, and variations of thermoinduced voltage, $\delta V_{10}$, are functions
of the heater temperature $T_h$.

In conclusion, the capacitive response of 2D electrons is determined not only by the thermodynamic
quantities $\partial n/\partial \mu$ and $\partial \mu/\partial T$ but also by the transport properties.
The recharging current Eq. (1) appearing as a response to weak periodic modulation of the voltage,
$V_0(t)=\Delta V \cos(\omega_0 t)$, is linear in the perturbation $\Delta V$ and determined by the
capacitance $C_{MW}$ which depends on thermodynamic density of states $\partial n/\partial \mu$ and
anomalous conductivity $\sigma_{\omega}$. The recharging current caused by a periodic
perturbation of heater temperature, $T_h(t)=T_{h0}+\Delta T_h \cos(\omega_0 t)$, depends on the
capacitance $C_{MW}$, entropy density $-\partial \mu/ \partial T$, and the thermoinduced voltage
$\delta V_{10}$. In the case of $\Delta T_h \ll T_h-T$, this current is linear in the
perturbation $\Delta T_h$.

Under conditions $X_k=2 \pi^2 k T/\hbar \omega_c \gg 1$, the thermal smearing of the Fermi
distribution leads to $\partial n/\partial \mu = \rho_{2D}$ and $\partial \mu/\partial T=0$.
The electron density of states cannot be probed by the capacitive response in this regime.
Nevertheless, the MW-induced features in the recharging current persist, and Eq. (24) is
rewritten in a simpler form
\begin{equation}
\delta Q=C_{0} \frac{\delta V_0  + \delta V_{10}(T_h)}{1+ (a_B/4d_G)/(1+\sigma_{\omega}/\sigma_{d}) },
\end{equation}
where $a_B=\hbar^2\epsilon_0/me^2$ is the Bohr radius. The quantity $\delta V_{10}$ is
also simplified in these conditions. Moreover, if we neglect the contribution of
magnetophonon oscillations [31], which is justified in the case of low temperatures [25]
and is applicable to our experiment since the amplitude of the oscillations is much
smaller than the background, $\delta V_{10}$ is written as
\begin{equation}
\delta V_{10}(T_h) \simeq  \frac{v_{12}(T_h)}{1+\sigma_{\omega}/\sigma_d} + v_{20}(T_h),
\end{equation}
where $v_{12}$ and $v_{20}$ are $B$-independent voltages describing classical phonon
drag contributions. In particular, under condition that $N_{\lambda {\bf Q}}({\bf r})
\propto {\bf q} \cdot \nabla T$, one has $v_{12}=-\alpha (T_1-T_2)$ and $v_{20}=
-\alpha (T_2-T_0)$, where $T_1$ is the temperature under the gate, $T_2$ and $T_0$ are
the temperatures in the points $(L/2,0)$ and $(L/2, L/2)$, respectively (see Fig. 5),
and $\alpha$ is the classical phonon-drag thermopower.

\begin{figure}[ht]
\includegraphics[width=9cm]{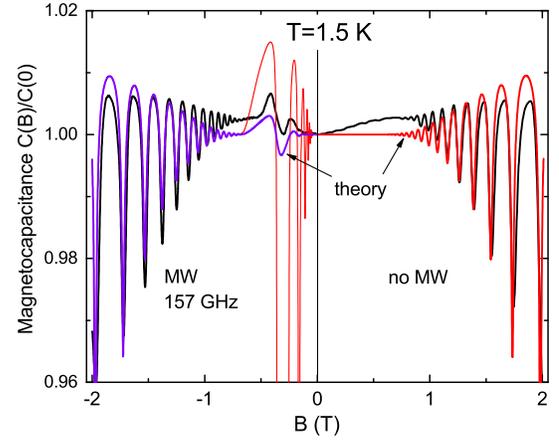}
\caption{(Color online) Comparison of calculated and experimental magnetocapacitance. The
right and the left parts show the dark and MW-modified capacitance, respectively. In the
left, two calculated plots are presented, the one given by the thicker line corresponds to
inhomogeneous suppression of anomalous conductivity: $\sigma_{\omega}$ is divided by
the factor $\xi$ defined by Eq. (27).}
\end{figure}

In spite of the smallness of the geometrical factor $a_B/4d_G \simeq 0.02$, the MW-modified
capacitance $C_{MW}$ is expected to change strongly compared to $C_0$ when the total
dissipative conductivity $\sigma_{tot}=\sigma_d+\sigma_{\omega}$ becomes small compared to
$\sigma_d$. Furthermore, for zero conductance ($\sigma_{tot}=0$) corresponding to the zero
resistance state realized in our samples, $C_{MW}$ formally goes to zero. However, we need
to emphasize that the theory given above cannot be applied in the intervals of magnetic
fields where the resistance is zero, because the distribution of currents and fields in
these conditions is far from the quasi-equilibrium one and is likely shaped into domain
structures [32].

The result of the calculation of the capacitance according to Eq. (20) is shown in
Fig. 6. The density of states needed for calculation of $\partial n/\partial \mu$,
$\sigma_{\omega}$, and $\sigma_d$ was described in the self-consistent Born approximation
based on the quantum lifetime of electrons of 7 ps, which is a typical value for our
samples at low temperatures.
The amplitude of fluctuations of chemical potential, $\Gamma=0.9$ meV, was found from
fitting the amplitudes of calculated and measured oscillations of the dark capacitance.
Weak oscillations of the chemical potential with magnetic field have been also taken
into account. The inelastic scattering time was estimated according to $\hbar/\tau_{in}
\simeq T^2/\mu$ [7]. The amplitude of the microwave field, $E_{\omega}$, was chosen in
such a way to obtain the closest correspondence between the measured and calculated
magnetoresistance. The theory describes the general behavior of the capacitance and
periodicity of the MW-induced oscillations. However, the amplitudes of the calculated
MW-induced oscillations are much larger than the observed ones. These deviations are
expectable from the theoretical point of view since the photovoltage associated with
the anomalous conductivity $\sigma_{\omega}$ develops on short length scales, near the
gate edge. Below we discuss this point in more detail.

We believe that the main reason for the small amplitudes of the observed oscillations is a
suppression of the influence of microwaves on 2D electron distribution in the presence
of spatial inhomogeneity near the gate edge. Indeed, while in the homogeneous case, the
energy relaxation of photoexcited electrons is determined by inelastic scattering only;
the relaxation of a spatially inhomogeneous distribution of electrons can also occur
via spatial diffusion. This effect is roughly described in terms of the enhancement
of the inelastic relaxation rate according to the substitution
\begin{equation}
\frac{1}{\tau_{in}} \rightarrow \frac{1}{\tau_{in}} + \frac{{\cal D}}{\ell^2}=
\frac{\xi}{\tau_{in}}, ~~ \xi=1 + \frac{l_{in}^2}{\ell^2},
\end{equation}
where $\ell$ is a characteristic scale of the inhomogeneity, ${\cal D}=v_F^2/2 \omega_c^2
\tau_{tr}$ is the diffusion coefficient ($v_F$ is the Fermi velocity), and $l_{in}=
\sqrt{{\cal D} \tau_{in}}$ is the inelastic diffusion length [33]. Since the anomalous
conductivity $\sigma_{\omega}$ is proportional to $\tau_{in}$, it will be suppressed as
a whole by the factor $\xi$. In the MIRO region of magnetic fields, $B \sim 0.2-0.5$ T,
$l_{in}$ is of the order of 1 $\mu$m. By choosing $\ell=0.22$ $\mu$m, we obtain a reasonably
good agreement between theory and experiment (see Fig. 6). The spatial inhomogeneities
of the potentials near the gate edge can appear because of gate fabrication imperfections.
Apart from this, screening of the MW field by the metallic gate can lead to a strong
spatial inhomogeneity of the MW intensity near the gate edge, which in turn contributes
to inhomogeneity of the electron distribution. Finally, one cannot exclude inhomogeneities
caused by nonlinear effects together with a dependence of the conductivities $\sigma_{\omega}$
and $\sigma_d$ on the applied voltage variations $\delta V_0$, though we did not see
direct indications of nonlinear behavior in our experiment. The suppression described by
Eq. (27) also explains why there is no indication of zero resistance states in the
capacitance signal.

We expect it is technically possible to reduce the suppression of the MW-induced
capacitance oscillations by increasing the distance to the gate, because an increase
in $d_G$ leads to increasing length $\ell$ and, consequently, to a reduction of
the factor $\xi$. The positive effect of this modification, however, would be partly
compensated by the overall decrease of the magnetocapacitance signal because of the
decrease in the geometric capacitance $C_0$.

A strong suppression of MIRO in spatially-inhomogeneous 2D electron gas was observed
previously in the samples patterned by antidot lattice with a period of the order of 1 $\mu$m [34].
This effect could not be attributed to lowering mobility in the patterned samples, as the
change in mobility compared to unpatterned samples was small. The suppression of MIRO
in Ref. 34 actually can be explained by the spatial diffusion mechanism described above.
Thus, the observation of the authors of Ref. 34 adds more confidence to the above
interpretation of our experimental results.

Now we proceed to discussion of the response to thermal modulation presented in Fig. 3.
In contrast to the case of voltage modulation, the effect of microwaves on the charging
current is strong and apparently not related to weak oscillations of the capacitance discussed
above. The theory suggests that the result of our observation can be understood as the
influence of microwaves on the magnetothermoelectric effect in Corbino geometry,
when the thermopower or, more generally, the thermoinduced voltage, is inversely proportional
to dissipative conductivity [28-30]. According to Eqs. (23) and (26), the MW-induced oscillations
of the conductivity contained in $\sigma_{\omega}$ are seen in the thermoinduced voltage as
inverted oscillations, where maxima are replaced by minima and vice versa. Indeed, by
applying the simple result of Eq. (26) with the ratio $v_{12}/v_{20}$ as a fitting parameter
to the interpretation of our experimental data, a reasonable agreement between the theory and
the experiment can be reached. The exception is the region of zero conductance, where the
theory is not applicable. The finite value of the charging voltage peak corresponding to
a zero resistance state can be explained by nonlinear effects which become increasingly
important when $\sigma_{tot}$ goes to zero or, even more likely, by transport anisotropy.
The latter means that the symmetry properties of thermoelectric tensor are modified,
and this can have a dramatic effect on thermoelectric phenomena under condition
of classically strong magnetic fields [35]. Indeed, we see a presence of antisymmetric
in $B$ contributions in the charging voltage signal, which can be a consequence of the
anisotropy. We leave this problem for further study. The influence of anisotropy on
thermopower in Corbino geometry has been already detected experimentally [28] but was
not studied systematically. We note that application of
MW irradiation allows us to study the region of relatively weak magnetic fields, whereas the
previous measurements of thermopower in Corbino geometry [28,30] demonstrated peculiarities
of thermoelectric response only in the quantum Hall regime.

\section{Summary and conclusions}

In this work, we explored the influence of microwaves on high-mobility 2D electron gas in a
GaAs quantum well layer partly covered by the gate. The capacitive coupling of electrons
to the gate placed in the center of the square sample allowed us to measure
the recharging current as a result of low-frequency modulation of the voltage between the
gate and one of the contacts. In this way, the effective MW-modified capacitance has
been measured directly. We also measured the recharging current as a response to periodic
variations of external heating by using the electrically driven heater placed under the
central part of the sample. Simultaneously, using Ohmic contacts, we measured the
magnetic-field dependence of the resistance that demonstrated the microwave-induced
resistance oscillations (MIRO) and zero resistance states.

We have found that in the presence of microwaves the capacitance oscillates with the magnetic
field. The period and the phase of these oscillations are the same as those of MIRO, but the
amplitude is much smaller. The phenomenon of microwave induced capacitance oscillations has
been explained as a result of violation of the Einstein relation between conductivity and the diffusion
coefficient under MW irradiation. This leads to a difference in the electrochemical potentials
of 2D electrons under the gate and out of the gate so that the relation between the induced
charge and the applied voltage [Eqs. (19) and (20)] is modified and becomes sensitive
to MW frequency. We note that a similar physical mechanism has been proposed [16] to describe
MW photovoltage oscillations [14,15] appearing under condition of a built-in electric field
near the contacts (for capacitive contacts, the photovoltage was measured as a response
to low-frequency modulation of microwave power). A comparison of the measured and
calculated MW-modified capacitance shows that the oscillations of the measured capacitance
have considerably smaller amplitudes. The discrepancy is lifted if one takes into
account a suppression of the influence of microwaves on electron distribution due to
spatial inhomogeneity of this distribution expected in the region near the gate edge.
Thus, we believe that our findings not only demonstrate the phenomenon of MW-induced
capacitance oscillations but also indicate the importance of spatial diffusion in
the relaxation of energy distribution of MW-excited electrons.

We have presented a linear response theory describing the recharging current and applicable
under the conditions of our experiment. In particular, the theory includes the effects of
inhomogeneous temperature and thermoinduced currents, which are missing in the previous
theoretical consideration [16,33,36] of the influence of MW irradiation on spatially-inhomogeneous
2D systems. The theory suggests that the response to a thermal perturbation depends on the
MW-modified capacitance described by Eq. (20). However, when the temperature gradient
and non-equilibrium phonon flux are present, the response has mostly thermoelectric origin.
Therefore, the main effect of the microwaves on the thermoinduced recharging current comes
from the influence of microwaves on the thermoelectric effect; the latter is likely dominated
by phonon drag. Whereas in our previous publication [17] we showed that MW irradiation
leads to antisymmetric in $B$ oscillating contribution of thermoelectric voltage, in the
present experiment we observe a symmetric in $B$ oscillating response in antiphase
with MIRO. This difference is actually related to the difference in the experimental
setups. In Ref. 17 we used Van der Pauw geometry with the heater placed outside the 2D layer,
while in this study we use Corbino-like geometry, where the gate at the sample center plays
the role of a capacitive contact, and the heating occurs from underneath the central
part of the sample. The absence of detailed knowledge about temperature and phonon flux
distribution in our samples does not allow a direct quantitative comparison of theory
and experiment at this point. Nevertheless, the periodicity and the shape of the oscillations
of the thermoinduced voltage clearly suggest that the signal reflects the inverse
proportionality of this voltage to the dissipative conductivity modified by microwaves.
In the regions of zero resistance states, the measured thermoinduced voltage shows high but
finite peaks. Since the linear theory developed for macroscopically homogeneous and isotropic
2D systems is not valid in these regions, a special consideration is required to describe
the height of the peaks.

Finally, we believe that our research shows that application of both capacitive
and thermoelectric methods in combination with MW irradiation may considerably extend
the possibilities for studies of 2D electron systems in high Landau levels. Further
research, both experimental and theoretical, is required to establish the role of
random inhomogeneities and anisotropy in transport and relaxation of non-equilibrium
2D electrons. \\

The financial support of this work by FAPESP, CNPq (Brazilian agencies) is acknowledged.
O.E.R. thanks I. A. Dmitriev for a helpful discussion.

\end{document}